\documentclass[]{cJAS2e}

\usepackage{subfigure,color}

\theoremstyle{plain}

\theoremstyle{remark}

\newcommand{\beginsupplement}{%
	\setcounter{table}{0}
	\renewcommand{\thetable}{S\arabic{table}}%
	\setcounter{figure}{0}
	\renewcommand{\thefigure}{S\arabic{figure}}%
}

\begin{document}

\jvol{00} \jnum{00} \jyear{2015} \jmonth{}


\title{Adaptive Trait Evolution in Random Environment}

\author{D.-C. Jhwueng$^{\rm a}$ $^{\ast}$\thanks{$^\ast$Corresponding author. Email: dcjhwueng@fcu.edu.tw
		\vspace{6pt}} and V. Maroulas$^{\rm b}$\\\vspace{6pt}  $^{a}${\em{Department of Statistics, Feng-Chia University, Taichung,Taiwan}}\\
	$^{b}${\em{Department of Mathematics, University of Tennessee, Knoxville, USA}}\\\received{v4.1 released March 2015} }

\maketitle

\begin{abstract} 
Current phylogenetic comparative methods generally employ the Ornstein-Uhlenbeck(OU) process for modeling trait evolution. Being able of tracking the optimum of a trait within a group of related species, the OU process provides information about the stabilizing selection where the population mean adopts a particular trait value. The optima of a trait may follow certain stochastic dynamics along the evolutionary history. In this paper, we extend the current framework by adopting a rate of evolution which behave according to pertinent stochastic dynamics. The novel model is applied 
to analyze about 225 datasets collected from the existing literature. Results validate that the new framework provides a better fit for the majority of these datasets. 


\begin{classcode}\textit{Classification codes}: 60H30, 62J12, 62P10\end{classcode}

\end{abstract}

\section{Introduction}
In evolutionary biology, phylogenetic comparative methods (PCMs) are commonly applied to analyze trait data for a group of species. Since species share evolutionary history, a good estimate of a phylogenetic tree, which represents the evolutionary relationship, is incorporated in data analysis. In past decades many comparative methods have been developed under different evolutionary hypotheses. For instance, the trait of a group of related species may rely on a continuous process ranging from either a single Brownian motion \cite{Felsenstein85}, or Brownian motions (BMv) \cite{OMeara06} to an Ornstein-Uhlenbeck (OU) process \cite{Hansen97, Butler04} or multiple optima, multiple rate of evolution and  multiple strength of selection OU process(OUmva) \cite{Beaulieu12}. For a more comprehensive review of comparative methods, the reader may refer to \cite{OMeara12}. 

Recently some PCMs have been developed for an advanced study of adaptive evolution in a randomly evolving environment \cite{Hansen08,Jhwueng14}. In contrast to the model of correlated evolution which sorely predicts the response trait, the model of adaptive evolution estimates the optimal relationship between two traits. To describe the model for adaptive evolution, we start with describing the evolutionary behavior for the response trait. 

Let $y_t$ be the trait value of a species at time $t$, $\theta_t$ be the optimum(evolutionary central tendency) of $y_t$, $\alpha_t^y$ be the rate of adaptation representing the speed of the trait tracking on its optimum, $\sigma_t^y$ be the rate of evolution of $y_t$, and $W_t^y$ be a white noise. A trait value of a species, $y_t,$ is a solution of the following Ornstein-Uhlenbeck (OU) stochastic differential equation (SDE)
\begin{equation} \label{OUevo1}
	dy_t=\alpha_t^y(\theta_t-y_t)dt +\sigma_t^y dW_t^y.
\end{equation}
Several models have been developed and applied widely to analyze trait data adopting further assumptions. For instance, the work in \cite{Hansen97} adopts equation \eqref{OUevo1} by assuming that the parameters are all time invariant that is $\alpha_t^y=\alpha_y, \theta_t^y=\theta_y$ and $\sigma_t^y=\sigma_y$ are all constant in equation (\ref{OUevo1}). The study in \cite{Butler04} assumes that multiple optima of $y_t$ occur during the evolutionary process. In this case, $\theta_t^y=\theta_\gamma$ where $\theta_\gamma$ is a piecewise constant value on the time interval $[t_{\gamma-1},t_{\gamma}]$ where $\gamma=1,2,\cdots,m$ and $t_0=0$ is the initial time of evolution and $t_m=T$ is the time length from $t=0$. Their model is then applied to study the evolution of the body size of \emph{anolis} lizard of the northern Lesser Antillean where each of these small islands supports either one or two species of anoles of different size (one species is large while the other is small). 
Beaulieu et al. \cite{Beaulieu12} extend the models considered in  \cite{Butler04} such that the force parameters, $\alpha_t=\alpha_\gamma$, and rate parameters, $\sigma_t=\sigma_\gamma$, are also allowed to take piecewise constant values on $[t_{\gamma-1},t_{\gamma}]$ and the models are applied to study the genome size evolution within a fairly large flowering plant clade. 

However, the optimum, $\theta_t$, typically is not a constant (or piecewise) static parameter. Instead its behavior typically evolves according to another independent OU process given below,
\begin{equation} \label{OUevo2}
	d\theta_t^y=-\alpha_t^\theta \theta_t^ydt + \sigma_t^\theta dW_t^\theta,
\end{equation}
where the parameters $\alpha_t^\theta, \sigma_t^\theta $ are the drift and the diffusion coefficients respectively for the optimum dynamics, and $W_t^\theta$ is a Brownian motion which could be correlated or independent from $W_t^y$ of eq. \eqref{OUevo1}. 
Based on these two OU evolutionary dynamics for the trait and its optimum, the model is called OUOU to reflect them. This model was established in \cite{Jhwueng14} in order to study the adaptation between body size and tail length of woodcreepers. A special case of the OUOU model is the work in  \cite{Hansen08} which considers that the trait follows the OU dynamics of equation \eqref{OUevo1}, and the optimum of the response trait, $\theta_t^y$, evolves via a Brownian motion, i.e. $\alpha_t^\theta=0$ and $\sigma_t^\theta$ is time invariant. The OUBM model in turn was applied to study whether the sexual size dimorphism increases with body size when the female is the smaller sex in primates. 

The rate of evolution $\sigma_t^y$  in equation (\ref{OUevo1}) measures the changing speed of the trait during its evolutionary process.  A trait $y_t$ described in equation (\ref{OUevo1}) may have a large variation when a species has evolved through the entire evolutionary history. Therefore this behavior cannot be captured by considering  a constant rate of evolution as it has been presumed in the current existing literature. Indeed, there are many traits from a group of related species with wide range of the evolutionary rate. For instance, Yopak et al. \cite{Yopak07} studied the variation in the brain organization of sharks. The widespread variation indicates the significant evolutionary diversity in their brain size and body mass. In such case, the variation should be captured in a stochastic way. Adopting these considerations, we develop and study in this paper the so-called OUBMBM and OUOUBM models, respectively, by treating $\sigma_t^y$ as a Brownian motion with a constant variance coefficient $\tau,$
\begin{equation} \label{OUevo3}
	d\sigma_t^y= \tau dW_t^\sigma,
\end{equation}
where the acronym OUBMBM (respectively OUOUBM) reflects the OU dynamics for the trait, a Brownian motion (respectively OU) for the optimum and a Brownian motion for the rate of evolution. Table 1 summarizes the models of trait evolution building on the general OU process that described by the SDE system of equation (\ref{OUevo4}).

\begin{table} 	\label{TraitModels}	
	\centering
	\begin{tabular}{cccccccc}
		\hline
		Model& $\alpha_t^y$ & $\alpha_t^\theta$ & $\theta_t^y$ & $\sigma_t^y$  & $\sigma_t^\theta$ & $\tau$  &Ref.\\ \hline
		
		BM & 0 & 0 & $\theta_y$ & $\sigma_y$  & $0$ & $0$  & \cite{Felsenstein85}\\ 
		
		OU& $\alpha_y$ & 0 & $\theta^y$ & $\sigma^y$  & $0$ & $0$  &\cite{Hansen97}\\ 
		
		OUm& $\alpha_y$ & $0$ & $\theta_{y,\gamma}$ & $\sigma_y$  & $0$ & $0$  &\cite{Butler04}\\
		
		BMv& $0$ & $0$ & $\theta_y$ & $\sigma_{y,\gamma}$  & $0$ & $0$  &\cite{OMeara06}\\ 
		
		OUmva& $\alpha_{y,\gamma}$ & 0 & $\theta_{y,\gamma}$ & $\sigma_{y,\gamma}$  & $\sigma_{\theta,\gamma}$ & $0$  &\cite{Beaulieu12}\\ 
		
		OUBM& $\alpha_y$ & $0$ & $\theta_t^y$ & $\sigma_y$  & $\sigma_\theta$ & $0$  &\cite{Hansen08}\\ 
		
		OUOU& $\alpha_y$ & $\alpha_\theta$ & $\theta_t^y$ & $\sigma_y$  & $\sigma_\theta$ & $0$  &\cite{Jhwueng14}\\ 
		
		OUBMBM & $\alpha_t^y$ & $0$ & $\theta_t^y$ & $\sigma_t^y$  & $\sigma_t^\theta$ & $\tau$  & \\ 
		
		OUOUBM& $\alpha_t^y$ & $\alpha_t^\theta$ & $\theta_t^y$ & $\sigma_t^y$  & $\sigma_t^\theta$ & $\tau$  &\\ \hline		
	\end{tabular}
	\caption{Models for trait evolution developed under the OU process. Parameters in the table were either assumed as of constant values (0 is included if the model does not use the parameter), piecewise constants value or random variable. For instance, the OU model in \cite{Hansen97} described the trait evolution using the typical Orsntein-Uhlenbeck process  while the OUmva model described the trait evolution under the generalized Hansen model \cite{Beaulieu12} with multiple optimum($\theta_\gamma$), multiple rate of evolution($\sigma_\gamma$) and multiple constraining forces($\alpha_\gamma$), $\gamma=1,2,\cdots,m$. The BMv model assumed trait evolved with multiple rate of evolution under the Brownian motion \cite{OMeara06}. }
\end{table}
Section 2 presents preliminary results for our models of adaptation evolution and gives a precise mathematical formulation. Section 3 considers the novel models and fits them to large datasets in literature and compare them using the coefficient of determination ($r$-squared value), the Akaike information criterion(AIC) and an assessment of the bias of parameters. Last, Section 4 offers a discussion along with concluding remarks of this paper. 

\section{Modeling adaptive evolution with random rate of evolution}
 
\begin{figure} 
	\vspace{-10pt}
	\begin{center}
		\includegraphics[width=14cm,height=8cm,angle=0]{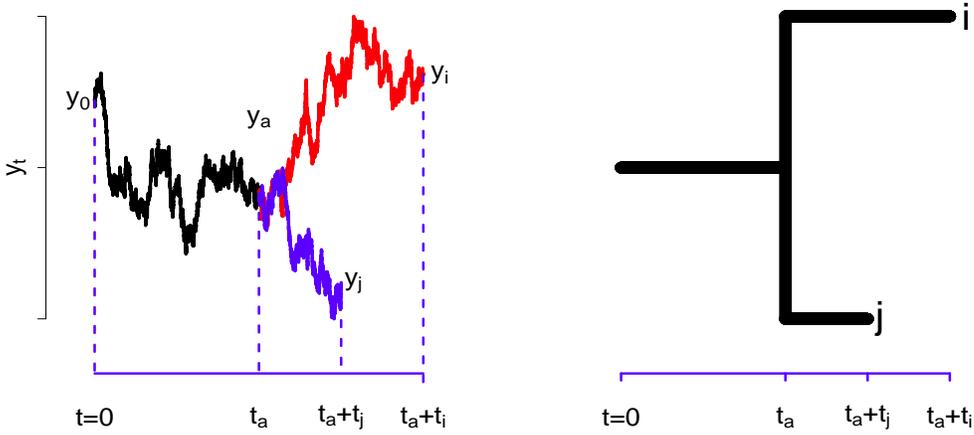}
	\end{center}
	\vspace{-15pt}
	\caption{An evolutionary tree for two species. The plot in the right panel shows the hypothetical evolutionary relationship between species $i$ and species $j$.  The plot in left panel represents the simulated trait evolution for the two species. Starting from time $t=0$ (with trait value $y_0$) to time $t=t_a$ (with trait value $y_a$), the two species share a common ancestor. Then they diverge and evolve independently for $t>t_a$ into two species $i, j$  at time $t=t_a+t_i$ and $t=t_a+t_j$ with trait value $y_i$ and $y_j$, respectively.}
	\label{fig2}
\end{figure}

Adopting equations (\ref{OUevo1})-(\ref{OUevo3}),  the trait evolution is organized in a vector form, $\textbf{Z}_t=(y_t,\theta_t^y,\sigma_t^y)'$, which satisfies the following stochastics differential equation (SDE),
\begin{equation} \label{OUevo4}
	d\textbf{Z}_t= \textbf{AZ}_tdt + \textbf{D}_td\textbf{W}_t,
\end{equation}
where the upper triangle matrix 
$\textbf{A}_t= \left( \begin{array}{ccc}
-\alpha_t^y&\alpha_t^y&0\\
0&-\alpha_t^\theta&0\\
0&0&0\\ \end{array} \right)$ consists of the parameters of selection strength; and 
$\textbf{C}_t=\textbf{D}_t\textbf{D}_t^T = \mbox{diag}((\sigma_t^y)^2,(\sigma_t^\theta)^2,\tau^2)$ 
represents the associated covariance matrix and $\textbf{W}_t=(W_t^y,W_t^\theta,W_t^\sigma)^T$ is the vector of the associated independent Brownian motions. Note that the different noises are assumed mutually independent.
We further assume that the force parameters are time invariant (i.e. $\textbf{A}_t=\textbf{A}$ is a constant matrix) and the rate of evolution for the optimum in equation (\ref{OUevo2}) is a constant (i.e. $\sigma^\theta_t=\sigma_\theta$). The SDE system described by Eq. (\ref{OUevo4}) has a unique solution 
\begin{equation} \label{OUevo5}
	\textbf{Z}_t=  e^{-\textbf{A}t} \textbf{Z}_0+ \int_0^t e^{-\textbf{A}(t-s)}\textbf{D}_s d\textbf{W}_s,
\end{equation}
where $\textbf{Z}_0=(y_0,\theta_0,\sigma_0^y)^T$ is the initial condition for $\textbf{Z}_t$ at $t=0$. The expected value of the solution $Z_t$, $\mathbb{E}[\textbf{Z}_t]= \textbf{Z}_0 e^{-\textbf{A}t}$ and  the second moment of the random vector $\textbf{Z}_t$, denoted by $\textbf{P}_t=\mathbb{E}[\textbf{Z}_t\textbf{Z}_t^T]$, can uniquely be determined by solving the system of an ordinary differential equation 
\begin{equation}
	\frac{d}{dt}\textbf{P}_t =\textbf{AP}_t+\textbf{P}_t\textbf{A}^T+ \mathbb{E}\textbf{C}_t. \label{Pt} 
\end{equation}
Next, we generalize the above model described by the SDE system in equation \eqref{OUevo4} for a group of $n$ interacting species which share an evolutionary history described by a phylogenetic tree $\Psi$. Let us consider that the current observed trait of the $i$th species, $y_{i,t}$, is a solution of the SDE given in equation (\ref{OUevo1}) where $0 \leq t \leq T$ and $T$ is the evolutionary time from the root of the tree to present time. Moreover, it is assumed that any pair of species, $(i,j)$ share a most recent common ancestor at time instant $t=t_a$ where $t_a \in [0,T]$ represents the evolutionary time from the root to the most recent common ancestor of the two species. To derive the joint distribution of the pair of random variables $(y_{i,t},y_{j,t}),  0 \leq t \leq T$, we need to incorporate the corresponding shape and length (evolutionary time) from the phylogenetic tree into the models described in equation (\ref{OUevo1}). Figure \ref{fig2} displays a cartoon of a hypothetical evolutionary relationship between species $i$ and species $j$ where the affinity between the two species scaled in time unit can be represented by the following matrix $\textbf{G}$ as shown in \cite{Jhwueng13}
\begin{equation}\label{simmtx}
\textbf{G}=\bordermatrix{
	&\text{species}~~i& \text{species}~~j   \cr
	\text{species}~~i&  t_a+t_i  & t_a      \cr
	\text{species}~~j&  t_a  &  t_a+t_j     \cr
}.
\end{equation}
Let us define that $E[y_i|y_a]$ is the expected trait value of species $i$ conditioned on its ancestral trait value $y_a,$ where $y_a$ is the trait value of the most recent common ancestor for species $i$ and species $j$. The associated covariance of a pair of traits $(y_i,y_j)$ for species $i$, $j$ that diverged at time $t_a$ and evolved independently thereafter is given by $Cov[y_i,y_j]=Cov[E[y_i|y_a],E[y_i|y_a]]$ \cite{Hansen96}. The study in \cite{Felsenstein85} considers a Brownian trait evolution which yields that the covariance of trait between species $i$ and species $j$ is proportional to the evolutionary time, that is $Cov[y_i,y_j]=\sigma_y^2t_a$.  When an Ornstein-Uhlenbeck process is considered for a single trait evolution, the manuscript \cite{Hansen96} establishes that the associated covariance, $Cov[y_i,y_j]=\sigma_y^2 e^{-2t_{ij}}/(2\alpha_y)$ when the initial condition $y_0$ is assumed to be a random variable. Furthermore, if conditioned $y_0$ on the root, then the associated covariance between two species trait evolved under the OU process is $Cov[y_i,y_j]=\sigma_y^2 e^{-t_{ij}}(1-e^{-2\alpha_y t_a})/(2\alpha)$ where the term  $t_{ij}$ denoted the evolutionary distance since the two species diverged and evolved independently (i.e. $t_{ij}=t_i+t_j$) \cite{Jhwueng14}.
When the optimum of the response trait evolves randomly, Hansen et al. \cite{Hansen08} and Jhwueng and Maroulas \cite{Jhwueng14} demonstrated that the variance-covariance structure between species $i$ and species $j$ under the OUBM model and OUOU model, respectively, can be derived under the following setting,   
\begin{equation} \label{OUBMBMcov}
	Cov[y_i,y_j]  =  c_{y_a}^2 Var[y_a] + c_{\theta_a}^2 Var[\theta_a] + 2c_{y_a}c_{\theta_a} Cov[y_a,\theta_a],
\end{equation} 
where $Var[y_a],Var[\theta_a]$ and $Cov[y_a,\theta_a]$ can be derived from $\textbf{P}_t$ by solving equation (\ref{Pt}) with initial condition at $t=t_a$. In particular, we have $c_{y_a}=e^{-\alpha_y t_{ij}/2}$ and $c_{\theta_a}=e^{-(\alpha_y-\alpha_\theta) t_{ij}/2}$ for OUBMBM model and $c_{y_a}=  e^{-\alpha_y t_{ij}/2}$ and $c_{\theta_a}=\frac{\alpha_y}{\alpha_y - \alpha_\theta} (e^{-\alpha_\theta t_{ij}/2} -e^{-\alpha_y t_{ij}/2})$ for OUOUBM model. We provide the derivation of $c_{y_a}$ and $c_{\theta_a}$ for the OUOUBM model in the Supplementary material. The  $c_{y_a}$ and $c_{\theta_a}$ for OUBMBM model can be derived in a similar manner.

Once the covariance between two species is determined, the next step is to develop the statistical model for regression analysis for studying the adaptive relationship among traits. First, we quantify the evolutionary relationship between the predictor, $x_t$, and the response trait, $y_t$. Typically, the optimum of the response trait, $\theta_t$, has a linear relationship with the predictor $x_t$ (i.e. $\theta_t=b_0+b_1x_t$). Note that a general functional relationship between $\theta_t$ and $x_t$ (i.e. $\theta_t=f(x_t)$) can indeed be the case and under this circumstance a linear approximation may be considered therein for parameter estimation. However, since the trait is typically transformed into the log scale before proceeding with the data analysis and the log transform converts nonlinear relationship into linear one, we use a linear relation in this work.  

Let $(x_i,y_i),i=1,2,\cdots,n$ be the pair of the trait values of species $i$ observed at the tip of the phylogeny. The joint distribution of the entire evolutionary history for these two traits can be modeled by a bivariate random variable $(x_{i,t},y_{i,t}), 0 \leq t \leq T$. Since $x_t$ is assumed to have a linear relationship with the optimum of $y_t$, the covariance for $(x_{i,t},y_{i,t})$ can be determined by $d\theta_t=b_1dx_t$.
The rate of evolution for the optimum, $\sigma_\theta$, is identified once $b_1$ and $\sigma_x^2$ are known( i.e. $\sigma^2_\theta=b_1\sigma_x^2$). The predicted evolutionary regression of $y$ on $x$ can be derived as $E[y|x]=k+\rho(t)b_1x$, where $k$ is a constant and $\rho(t)=Cov[y_t,\theta_t]/Var[\theta_t]$, see e.g. \cite{Jhwueng14}. 

If the response $Y=(y_1,y_2,\cdots,y_n)$ and the predictor $X=(x_1,x_2,\cdots,x_n)^T$ are observed at the tips of the phylogeny with the assumption that the relationship of the primary optimum $\theta_t$ to the predictor variable is a simple linear regression, the evolutionary regression curve for $(X,Y)$ has the form
\begin{equation} \label{RegressionModel}
	Y=\textbf{X}\beta+r
\end{equation}
where $\textbf{X}=[\textbf{1},\rho(t)X_1,\cdots,\rho(t)X_q]$ is the designed matrix of size $b \times q$, \textbf{1} is the vector of ones, $b=(b_0,b_1,\cdots,b_{q-1})^T$ is a $q$ dimensional vector of regression parameters, $r$ is the residual vector following a normal distribution with zero mean vector, and with the residual covariance matrix $\textbf{V}$ given below
\begin{equation} \label{ResidualCov}
	\textbf{V}_{ij}=Cov[r_i,r_j]=Cov[y_i-E[y_i|\theta_i], y_j-E[y_j|\theta_j]]
\end{equation}
where $E[y_t|\theta_t]=\hat{\beta}_0(t)+\hat{\beta}_1(t) \theta_t$ is the regression of the trait on the optimum (see Lemma 1 in \cite{Jhwueng14} for the derivation under the OUOU model, the optimal regression for other models described here can be derived in a similar manner). Note that equation (\ref{ResidualCov}) involved four terms, $Cov[y_i,y_j], Cov[y_i,E[y_j|\theta_j]], Cov[y_j,E[y_i|\theta_i]]$ and $Cov[E[y_i|\theta_i],E[y_j|\theta_j]]$ which are computed once  $\mathbf{\Sigma}_t$ was determined by solving equation (\ref{Pt}) with initial condition $\textbf{Z}_0$. 

The statistical model derived from the OU dynamic of evolution is a multivariate normal distribution(i.e. $Y \sim \text{MVN} (E[Xb], \textbf{V})$). The log likelihood for the regression analysis is 
\begin{equation} \label{LikeFcn}
	\log L(b,\textbf{V}|X,Y,\Psi )= \log \left[\frac{1}{\sqrt{ (2\pi)^n \det (\textbf{V})}} e^{-\frac{1}{2}(Y-Xb)^t\textbf{V}^{-1}(Y-Xb)} \right]
\end{equation}
where $\Psi$ is a rooted phylogenetic tree with known topology and branch lengths and can be transformed to the matrix $\textbf{G}$ directly for further use of constructing the matrix $\textbf{V}$. The latter is computed for the different models using the open source SAGE \cite{Stein14}. For parameter estimation, we use a similar algorithm as in \cite{Jhwueng14} as follows.

Given the trait data $Y$ and phylogenetic tree, the algorithm starts the search with an ordinary lease square estimate $\hat{b}_0=(\textbf{X}'\textbf{X})^{-1}\textbf{X}'Y$. As $\Psi$ is given, the variance covariance structure for the residual $\textbf{V}$ is calculated using equation (\ref{ResidualCov}). The likelihood in equation (\ref{LikeFcn}) is then optimized and the MLEs are recorded. The regression estimates $\hat{b}$ is then updated by $\hat{b}=(\hat{\textbf{X}}'\hat{\textbf{V}}^{-1}\hat{\textbf{X}})^{-1}\hat{\textbf{X}}'\hat{\textbf{V}}^{-1}\textbf{Y}$.  This procedure is repeated until a distance between the updated regression estimate and previous estimate is within an upper bound. We use the \textit{R} packages \textit{optim} \cite{Nash14} for optimizing the likelihood. To address the potential sensitivity of the algorithm in the starting point, we use an alternative search for the MLEs where at most five different starting points are randomly selected in the domain of the parameter space and within each search,  at most five attempts are set to access the convergence of estimation. If the convergence is not detected after the maximum iteration is reached, a different starting point will be either chosen using the current estimate or randomly selected in the domain. We keep updating the searches whenever a new improvement of the likelihood is observed. The search stops and save the estimates when either three improvement of likelihood are found or the maximum number of searches are reached.


%
%
%
%
%

\section{Empirical study and simulations}

\subsection{Motivation}
\begin{figure} 	\label{fig1}
	\vspace{-10pt}
	\begin{center}
		\includegraphics[width=12cm,height=8cm,angle=0]{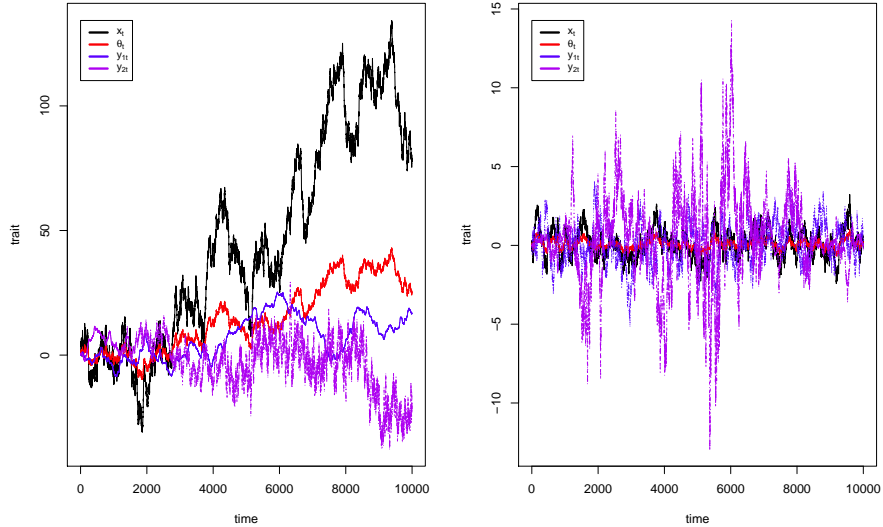}
	\end{center}
	\vspace{-15pt}
	\caption{
	Evolutionary paths for the trait models. The path for the predictor trait $x_t$(in black color) was first simulated under the Brownian motion or OU process; and the optimum paths $\theta_t$ (in red color) was then simulated using the linear relationship $\theta_t=b_0+b_1x_t$; finally, the path for the response traits were simulated. In the left panel, $y_{1t}$ represented path(in blue) simulated under the OUBM model, and $y_{2t}$ represented path(in purple) simulated under the OUBMBM; and in the right panel $y_{1t}$ represented path (in blue) simulated under the OUOU model, and $y_{2t}$ represented the path(in purple) under the OUOUBM model.}
\end{figure}

Before any actual data set is analyzed using the models developed herein and their comparison with the current literature, we start by simulating the evolutionary path for the models. The paths were generated using parameter values $(\alpha_y,\sigma_y,\sigma_\theta,b_0,b_1)=(0.05,0.01,0.32,0.50,0.32)$ for the OUBM model; $(\alpha_y,\sigma_\theta,b_0,b_1,\tau)=(0.05,0.32,0.50,0.32,0.01)$ for the OUBMBM model, $(\alpha_y,\alpha_\theta,\sigma_y,\sigma_\theta,b_0,b_1)=(0.01,0.01,0.20,0.45,0.05,0.30)$ for the OUOU model; and $(\alpha_y,\alpha_\theta,\sigma_y,\sigma_\theta,b_0,b_1,\tau)=(0.01,0.01,0.20,0.45,0.05,0.30,0.01)$ for the OUOUBM model. Figure \ref{fig1} shows simulation paths under the OUBM/OUBMBM models (the left panel) and the OUOU/OUOUBM models(the right panel).
 We calculate the standard deviation(sd) of the trait considering the evolution paths from time $t=0$ to $t=10000$ in this simulation for each model. In this simulation, the OUBM model has sd of value $72.64$ while the sd is $107.96$ for OUBMBM; The OUOU model has sd $1.61$ while the sd is $8.00$ for OUOUBM. The models OUBMBM and OUOUBM with the random rate of evolution account more variation than the special case models OUBM and OUOU, respectively which implies that evolutionary rate with high volatility should adopt a dynamic rate of evolution that is either an OUOUBM or OUBMBM model. 

\subsection{Shark dataset}
\begin{figure} 
	\begin{center}
		\includegraphics[width=14.5cm,height=15cm,angle=0]{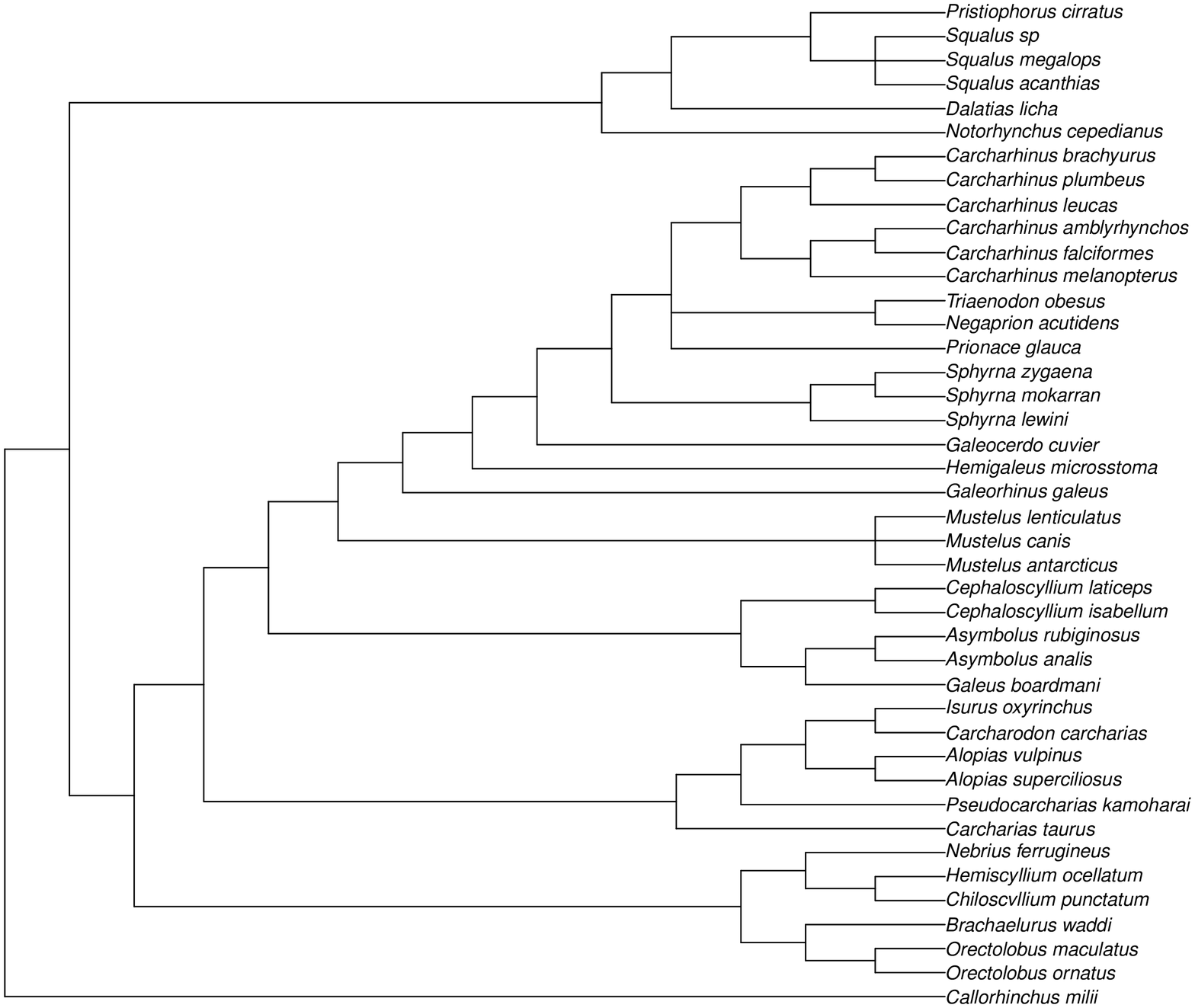}
	\end{center}
	\caption{Evolutionary relationship of the sharks replotted from \cite{Yopak07}.}
	\label{fig_sharkphylo}
\end{figure}

\begin{figure} 
	\begin{center}
		\includegraphics[width=14.5cm,height=15cm,angle=0]{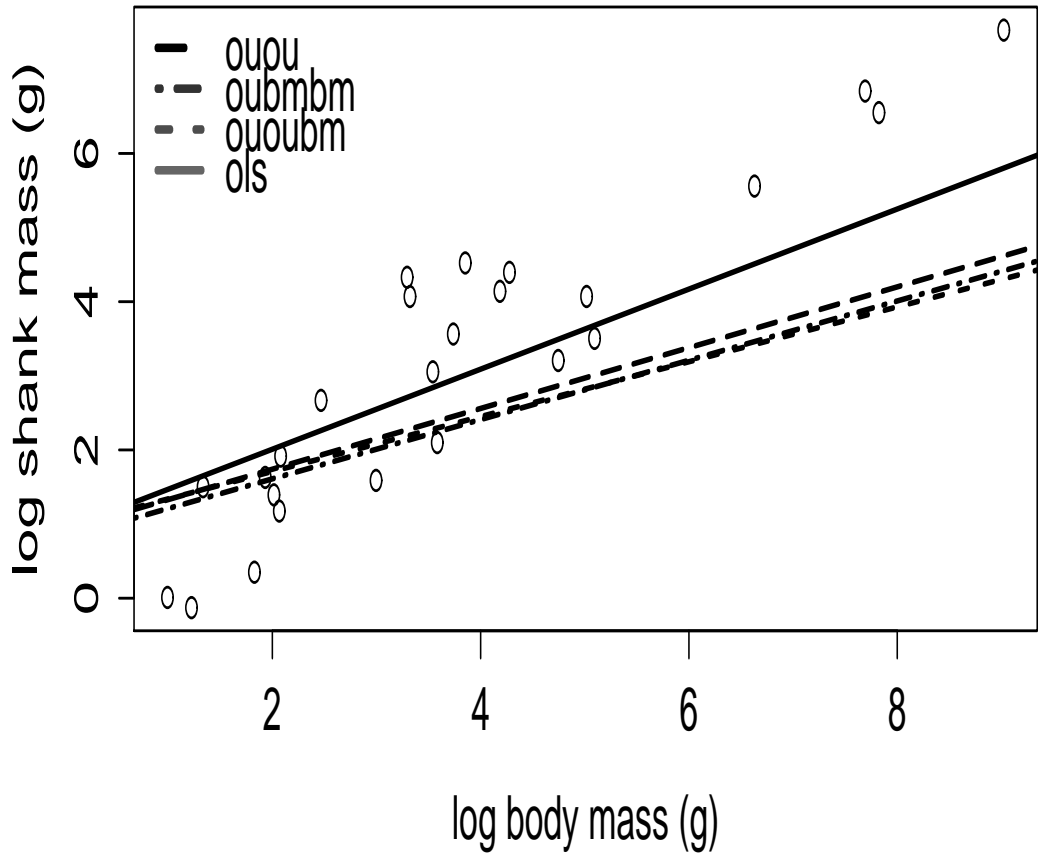}
	\end{center}
	\caption{Evolutionary regression curves.}
	\label{fig_YopakRegLine_ououbm}
\end{figure}

We first apply the proposed models herein that is the OUBMBM and OUOUBM as well as the OUOU model, which was suggested in \cite{Jhwueng14} as the general version of \cite{Hansen08}, in order to analyze the shark (\textit{chondrichthyans}) dataset in \cite{Yopak07}. A correlation study is conducted through the ordinary regression analysis and independent contrast method in \cite{Felsenstein85} and a significant relationship is found between the body size and brain mass. Figure \ref{fig_sharkphylo} shows the evolutionary relationship represented by a rooted phylogenetic tree among the 42 species of study interest. Due to high volatility in brain size (response variable) and the body weight (predictor variable), both datasets were log-transformed prior to data analysis.  Figure \ref{fig_YopakRegLine_ououbm} and Table \ref{EvoLinesTable} shows the regression results.

\begin{table} 	\label{EvoLinesTable}	
	\centering
	  
	\begin{tabular}{cccccccc}
		\hline
		Model& Regression Line & $r^2$ value  \\ \hline
		OUOU& $y=0.92+0.41x$ & $70\%$         \\
		OUBMBM& $y=0.81+0.40x$ & $72\%$       \\ 
		OUOUBM& $y=0.97+0.37x$ & $75\%$       \\ 
		 \hline		
	\end{tabular} \caption{Evolutionary regression lines}
\end{table} 

Comparing to the ordinary regression analysis(OLS) where $y=0.93+0.54x$ with $r^2=74\%$, we find that the regression slopes for the four models were slightly shallower than the slope using the OLS approach in this dataset. On the other hand, the regression slopes for the models deviate from 0, therefore our models support the conclusion that the relative brain development reflects the dimensionality of the environment prey caption in addition to phylogeny in adaptation aspect as well as using the OLS and independent contrast method \cite{Yopak07}. In addition, we find that for this wide spread data the general OUOUBM model ($r$-squared value $75\%$) and the OUBMBM model ($r$-squared value $72\%$) provides better fit than the OUOU model($r$-squared value $70\%$). These results are summarized in Table \ref{EvoLinesTable}.

\subsection{Models comparison}
\subsubsection{Accessing the statistical fit of the models via $r$-squared values}
\begin{figure}[htb]
	\centering
	\begin{tabular}{@{}cccc@{}}
		\includegraphics[width=.5\textwidth]{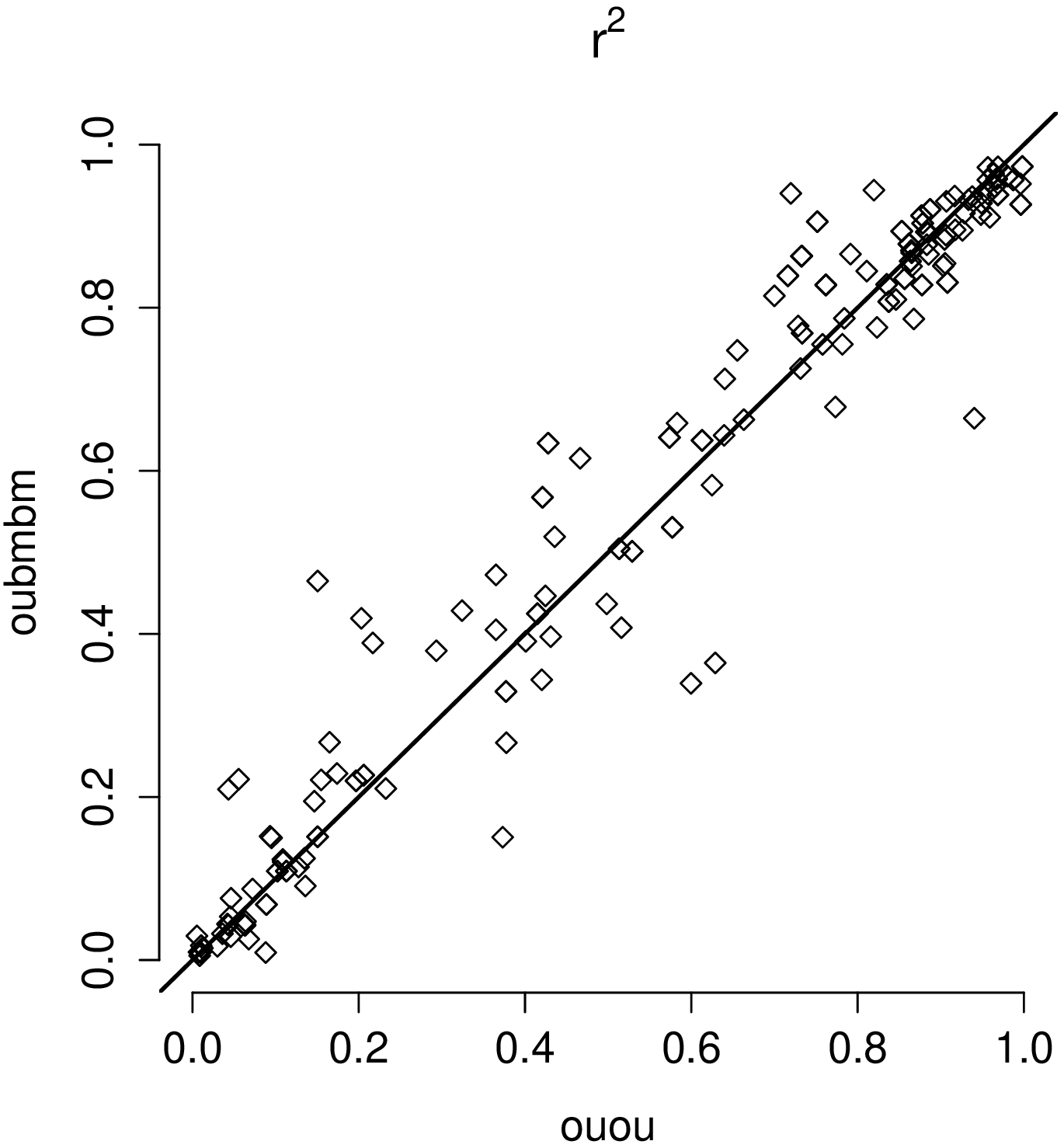} &
		\includegraphics[width=.5\textwidth]{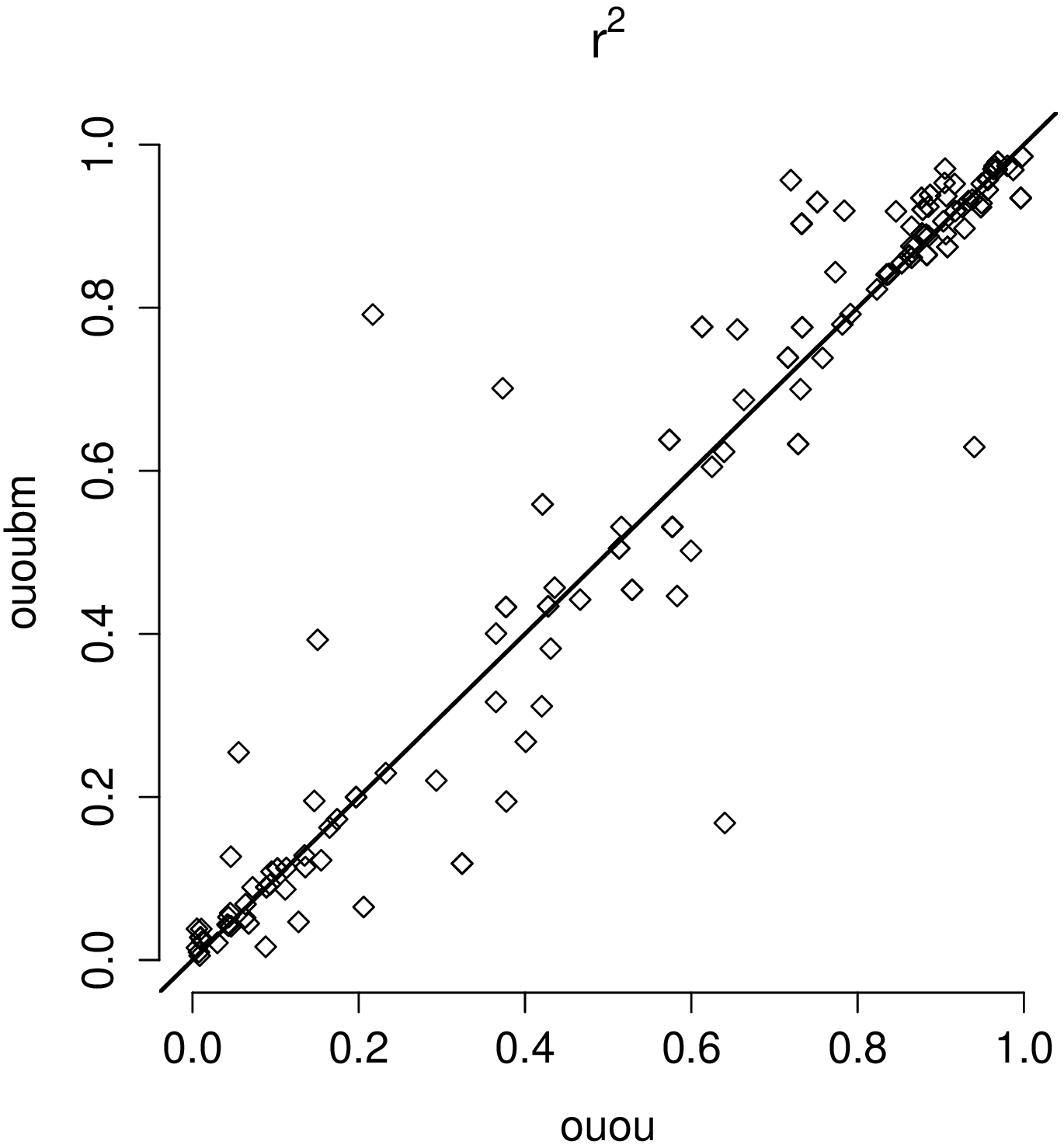} &
		  \\
	\end{tabular}
	\caption{Comparison of the models using $r$-squared values.}
	\label{fig_r_sq}
\end{figure}

In this Section, we consider 225 bivariate datasets appeared in the existing literature \cite{Aguirre02,Bonnie05,Melville03,Monnet02,Moreteau03,Niewiarowski04,Sanchez03,Tubaro02,Vanhooydonck02,Weiblen04} and we compare the performance and fitting of the OUBMBM and OUOUBM models with the general OUOU model \cite{Jhwueng14}. First, we summarize the comparison of the fit of these models using the coefficient of determination ($r$-squared values). The output of the analysis is given in Figure \ref{fig_r_sq} where each point in the plot represents the $r$-squared value for the models of our interest. We compare the OUOU model with the OUBMBM model and summarize the result on the left panel. The comparison between OUOU model and OUOUBM model is given on the right panel. In each plot, the 1:1 line shows the equivalent fit of the both models; points below the 1:1 line indicate a better fit of the model whose $r$-squared is in the horizontal axis (OUOU model) and a point above the 1:1 line yields a better fit for the model that has $r$-squared in the vertical axis(the OUBMBM model and OUOUBM model). Overall both figures indicate that the fit assessed by r-squared is consistent between the OUOU model and the new models (OUBMBM or OUOUBM) in most datasets. Most points are closed the diagonal line which indicates that  when a high/low $r-$squared is observed by the OUOU model, the new model can identify a similar $r-$squared as well.  Moreover, the OUBMBM model fits better than the OUOU model in 49.21 $\%$. This shows that there might be no significant difference for two models as they have a samilar ability to detect the fit.  On the other hand, the OUOUBM models proposed herein fits better than the existing OUOU model more frequently, in fact 57.40 $\%$ while the OUOU model fits better than the OUOUBM model for 42.60 $\%$ which accounts for a 14.80 $\%$ difference. 

\subsubsection{Accessing the statistical fit of the models via the Akaike Information Criterion}
\begin{figure}
	\begin{center}
		\includegraphics[width=15cm,height=15cm]{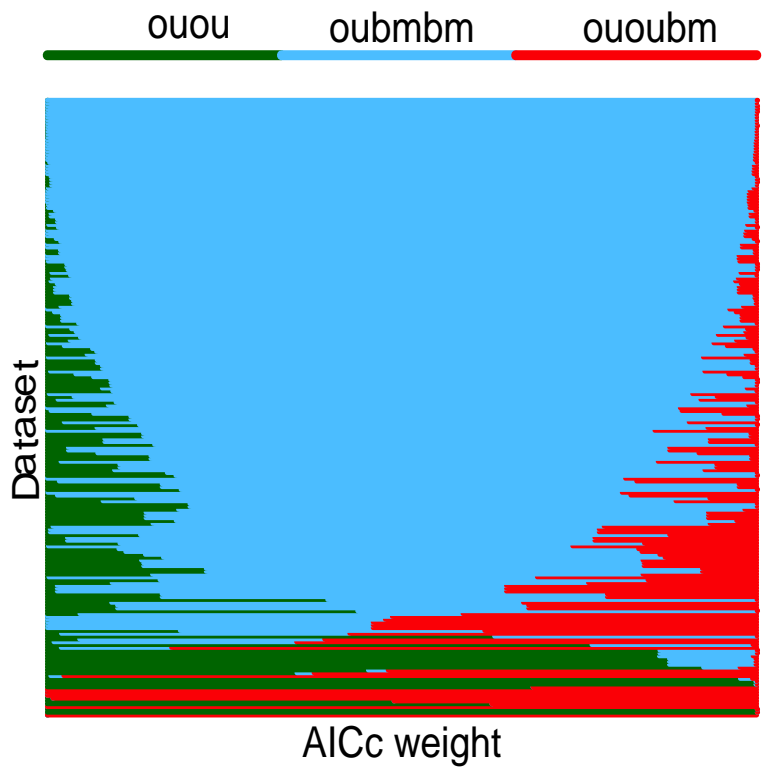}
	\end{center}
	\caption{The relative support measured by AICc weight among the three models in this study (OUOU, OUBMBM, and OUOUBM) across 225 datasets where each horizontal line represents a weight of the models. The OUBMBM model accounts for more support for a majority of the datasets.}
	 \label{fig_AICcWeight}
\end{figure}

Next we compare the models employing their corresponding AICc values \citep{Burham02} where  $AICc= -2\log L + 2nk / (n-k+1),$ $\log L$ is the associated log-likelihood value, $k$ is the number of model parameters and $n$ is the number of extant species. We display the result in Figure \ref{fig_AICcWeight} where the relative support for the models are reported by their corresponding Akaike weight $\omega=\exp(-0.5 \Delta AICc)$ where $\Delta AICc$ is the difference between the AICc value of the model and the minimal AICc value among the model set. Each horizontal line in Figure \ref{fig_AICcWeight} represents the scaled Akaike weights of the models. 

For most datasets, we found that OUBMBM model accounts for more Akaike weight and happens to be the AICc selected models when compared to the OUOU model and the OUOUBM model. Since the OUOU model and the OUOUBM model only differ at most in one parameter from the OUBMBM model, this result can be resulted from the likelihood values where the OUBMBM contributes to lower likelihood than the OUOU/OUOUBM. 

\subsubsection{Bias of parameters for the models}

We further access the bias of parameters for the models of interest.  We also include the OUBM model in this section. We considered to generate data with four different sample sizes of $n=16,32,64$, and $128$ using the parameters values of $(\alpha_y,\sigma_y)=(0.05,0.10)$ for OUBM, $(\alpha_y,\alpha_\theta,\sigma_y)=(0.05,0.12,0.10)$ for OUOU, $(\alpha_y,\tau)=(0.05,0.30)$ for OUBMBM and $(\alpha_y,\alpha_\theta,\tau)=(0.05,0.12,0.30)$ for OUOUBM.  The true regression estimate $(b_0,b_1)$ are set to $(1.20,0.72)$, and $50$ replicates are sampled from this set up with considering of using four different types of tree phylogeny  (1) a star tree where all species are completely unrelated, (2) a completely balanced tree, (3) a completely pectinate tree, and (4) a random tree generate under the birth-death process. 
Hence $4 \times 4 \times 4 \times 50 = 800$ replicates are generated and the associated parameters are estimated through the MLE approach described in Section 2. For each parameter of interest, we report the boxplots of parameter estimates $\alpha_y, \alpha_\theta, \sigma_y$, and $\tau$ under each models and different sample sizes (we combine the tree topology  effect in this study). The results are shown in Figure S1-S4 (see supplemental material section).  Overall, we find that the accuracy for estimating the parameters of interests is improved with increasing the sample size. The interquartile range(IQR) shrinks with larger sample size increases for each model.

Figure S1 and Figure S2 shows that estimation for $\alpha_y$ and $\alpha_\theta$, respectively, cannot achieve satisfactory accuracy because there are many outliers configured in the corresponding boxplots. Focusing on Figure S1, the parameter $\alpha_y$ is in general poorly estimated under the OUBM and the OUBMBM model. This situation also occurs for estimating $\alpha_\theta$ as in Figure S2 the parameter estimates become less spread as the sample size increases. Figure S3 and Figure S4 show that parameter estimation for $\sigma_y$ and $\tau$, respectively, are in general good. In Figure S3, OUOU model accounts for larger variation in estimating $\sigma_y$ than the OUBM model. Furthermore, $\sigma_y$ for the OUOU model can be estimated more accurately as the sample size increases. In Figure S4, the OUOUBM model provides wider IQR than the OUBMBM model for estimating the parameter $\tau$. We leave a discussion for this in the next session. 
\section{Conclusion}
In this paper, we developed two models for the adaptive trait evolution and evaluated their performance by analyzing many empirical datasets. We found that  our model  OUBMBM/OUOUBM fit better for more datasets than their submodel model OUOU when evaluated the performance of fit by the $r$-squared values. On the other side, under the likelihood based model selection criterion, we found that due to the likelihood, the OUBMBM model became AICc selected models for most empirical datasets than the OUOU and OUOUBM models. 

In bias study of parameter, we found that the parameters can not be always estimated accurately under the MLE analysis for many datasets. Ho and An\'{e} \cite{Ho13} and An\'{e} \cite{Ane08} described the limitation of the parameter estimation of the BM model described in \cite{Felsenstein85} and the limitation of the OU model described in \cite{Hansen97} for trait evolution. They pointed out that since some parameters may not identifiable, the maximum likelihood estimators for trait models could fail to be consistent estimators where the convergence to the true parameter cannot be guaranteed. For the OU model, Ho and An\'{e} \cite{Ho13} proved that the selection optimum cannot be estimated consistently as the tree grows indefinitely. In our framework, we might encounter the same problem when estimating the parameters for the models of adaptive evolution developed here. To deal with this situation,  we suggest a future study for the models of adaptive evolution under a Bayesian paradigm to reduce the estimation difficulty for non-identifiable parameters (see \cite{Uyeda14, Vill12} for works of comparative methods using a Bayesian approach). 

The simulation paths in Figure \ref{fig1} indicated the higher variation of the OUBMBM model with respect to the OUBM model, and the higher variation of the OUOUBM model with respect to the OUOU model. It is expected that the OUBMBM/OUOUBM model would provide a better fit than the OUOU model for data with wider variation. In the shark dataset, we demonstrate that the both OUBMBM and OUOUBM models provide a better fit than the OUOU model.    


The SAGE source code for solving the moments of $\textbf{Z}_t$ and the covariance between the residual $\textbf{V}$, the R source codes for parameter estimation and simulations and the datasets used in this work can be accessed directly at www.tonyjhwueng.info/OUOUBMsim.


\subsection{Acknowledgements}


\subsection{Funding}




\subsection{Supplemental material}
 \beginsupplement
 \subsubsection{Variance-covariance structure for the OUOUBM model}
 
 For OUOUBM model, $A= \left( \begin{array}{ccc}
 -\alpha_y&\alpha_y&0\\
 0&-\alpha_\theta&0\\
 0&0&0\\ \end{array} \right)$. Solving equation (\ref{OUevo5}) amounts to calculate the exponential $A$. Note that $A$ has three distinct eigenvalues $\{-\alpha_y,-\alpha_\theta,0\}$ with eigenvectors set $ \textbf{Q}= \{ (1,0,0)',(\frac{\alpha_y}{\alpha_y-\alpha_\theta},1,0)',(0,0,1)'\}$.  This yields that the term $e^{\textbf{A}t}$ can be computed straightforwardly by $e^{-At}=\textbf{Q} \Lambda \textbf{Q}^{-1}$ where $\Lambda=\text{diag}\{ e^{-\alpha_yt},e^{-\alpha_\theta t}, 1 \}$ is the diagonal matrix and $\textbf{Q}=\{v_1,v_2,v_3\}$ is the matrix of eigenvectors. Since the white noises $\textbf{W}_t$ have expectation zero, the expected value of $\textbf{Z}_t$ can be determined given the initial condition $\textbf{Z}_0=(y_0,\theta_0,\sigma_0)'$. In particular, the expecated value of $y_t$ conditioned on the initial value $y(0)=y_0$ has the form 
   \begin{align}
  {E}[y_t|y_0] &=c_{y_t} y_0 + c_{\theta_t} \theta_0  \label{expect_y_0}
  \end{align}
  where $c_{y_t}=e^{-\alpha_yt}$ and $c_{\theta_t}=\frac{\alpha_y}{\alpha_y-\alpha_\theta}\left(e^{-\alpha_\theta t}-e^{-\alpha_y t}\right)$.
  
Then we can apply equation (\ref{expect_y_0}) to $y_t$ conditioned on any ancetral value, that is $ {E}[y_t|y_a]= c_{y_t} y_a + c_{\theta_t} \theta_a$ where $y_a$ and $\theta_a$ are the ancetral value at $t=t_a$. In \cite{Hansen96}, the associated covariance of a pair of traits $(y_i,y_j)$ for species $i$, $j$ that diverged at time $t_a$ and evolved independently thereafter is given by $Cov[y_i,y_j]=Cov[E[y_i|y_a],E[y_i|y_a]]$  where $E[y_i|y_a] = e^{-\alpha_y t_ij/2} y_a + \frac{\alpha_y}{\alpha_y-\alpha_\theta}\left(e^{-\alpha_\theta t_{ij}/2}-e^{-\alpha_y t_{ij}/2}\right)\theta_a$ and $t_{ij}$ is the evolutionary distance since species $i$ and $i$ diverged. 
 
To complete the calculation of $Cov[y_i,y_j]$, the next step is to calculate the terms $Cov(y_a,y_a), Cov(y_a,\theta_a)$ and $Cov(\theta_a,\theta_a)$. These term can be determined by solving the ordinary differential equations (ODEs) in equation (\ref{Pt}). Notice that although there are $3 \times 3 = 9$ ODEs in  Eq. (\ref{Pt}), due to symmetry of $\textbf{P}_t$ it suffices to solve six equations ( including the three equations for the second moments of $y_t,\theta_t, \sigma_t$) and the expected value of $y_t\theta_t, y_t\sigma_t$ and $\theta_t\sigma_t$).  Since some variables are embbeded in the equations, we cannot solve the six ODEs simultaneously. Fortunately, we can determine a solution recursively once upon one ODE is completely solved and it is then used for solving another ODE.

\begin{figure}
	\begin{center}
		\includegraphics[width=15cm,height=15cm]{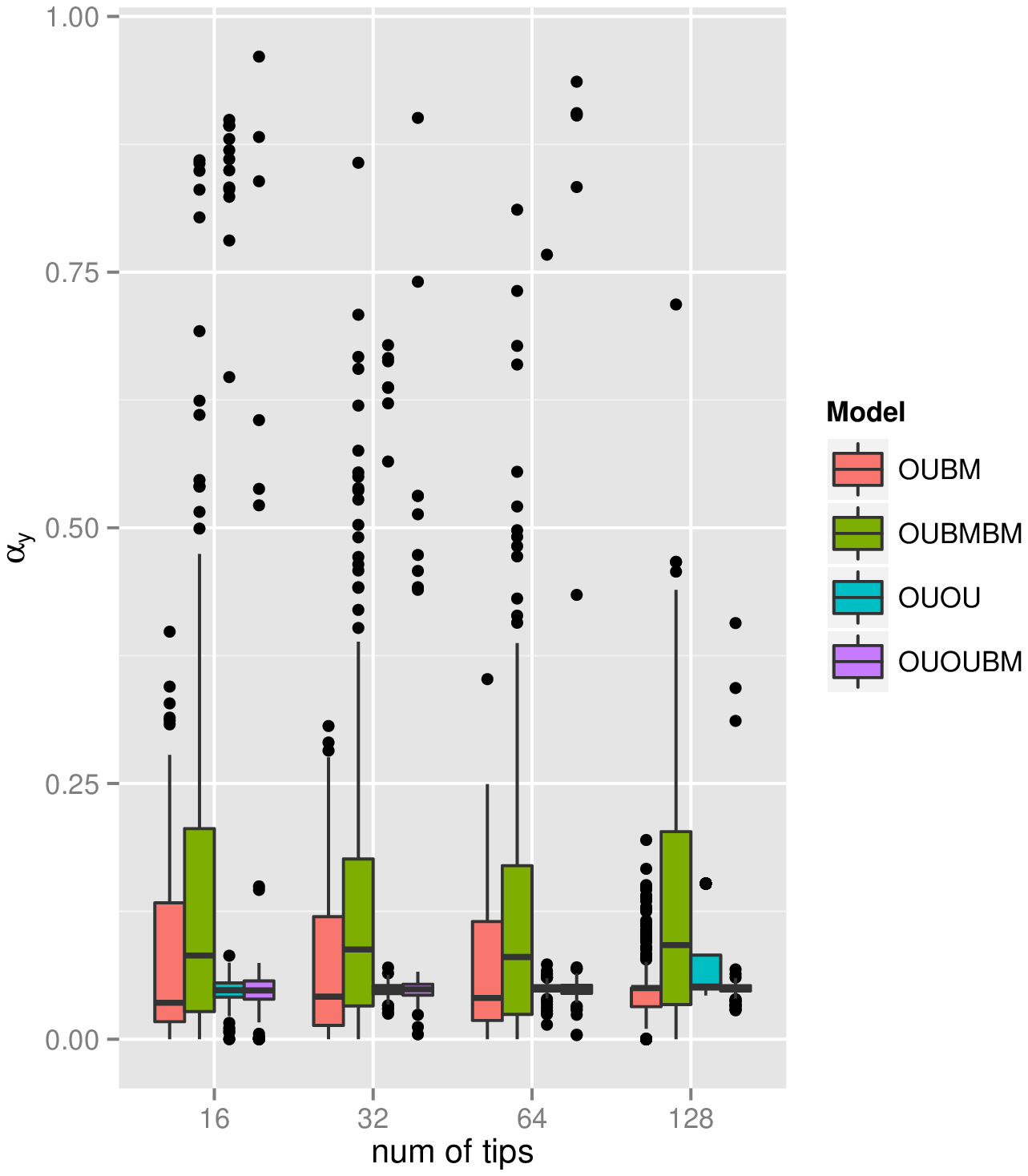}
	\end{center}
	\caption{Boxplot for $\alpha_y$.}
	\label{alpha_y}
\end{figure}

\begin{figure}
	\begin{center}
		\includegraphics[width=15cm,height=15cm]{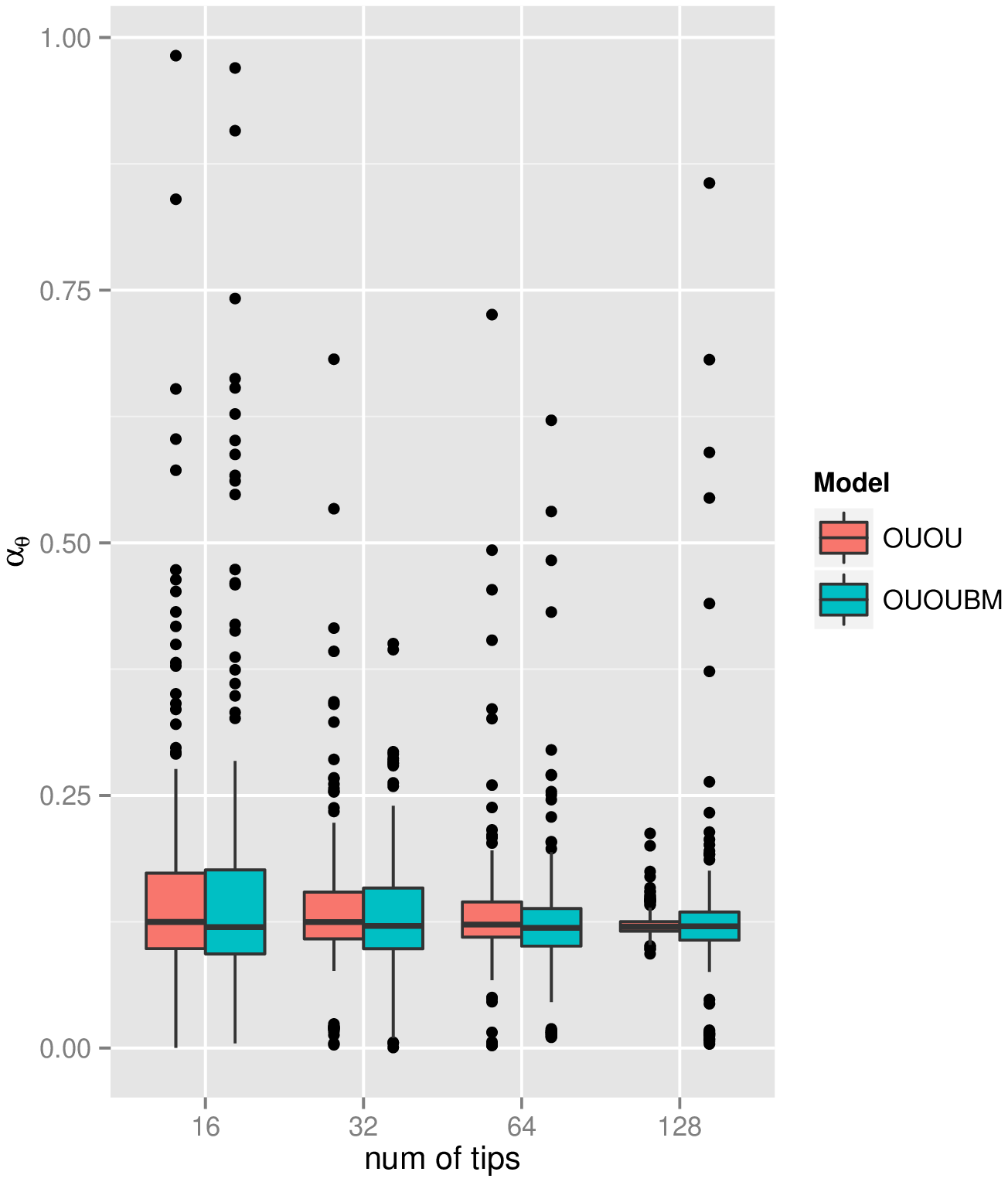}
	\end{center}
	\caption{Boxplot for $\alpha_\theta$.}
	\label{alpha_theta}
\end{figure}

\begin{figure}
	\begin{center}
		\includegraphics[width=15cm,height=15cm]{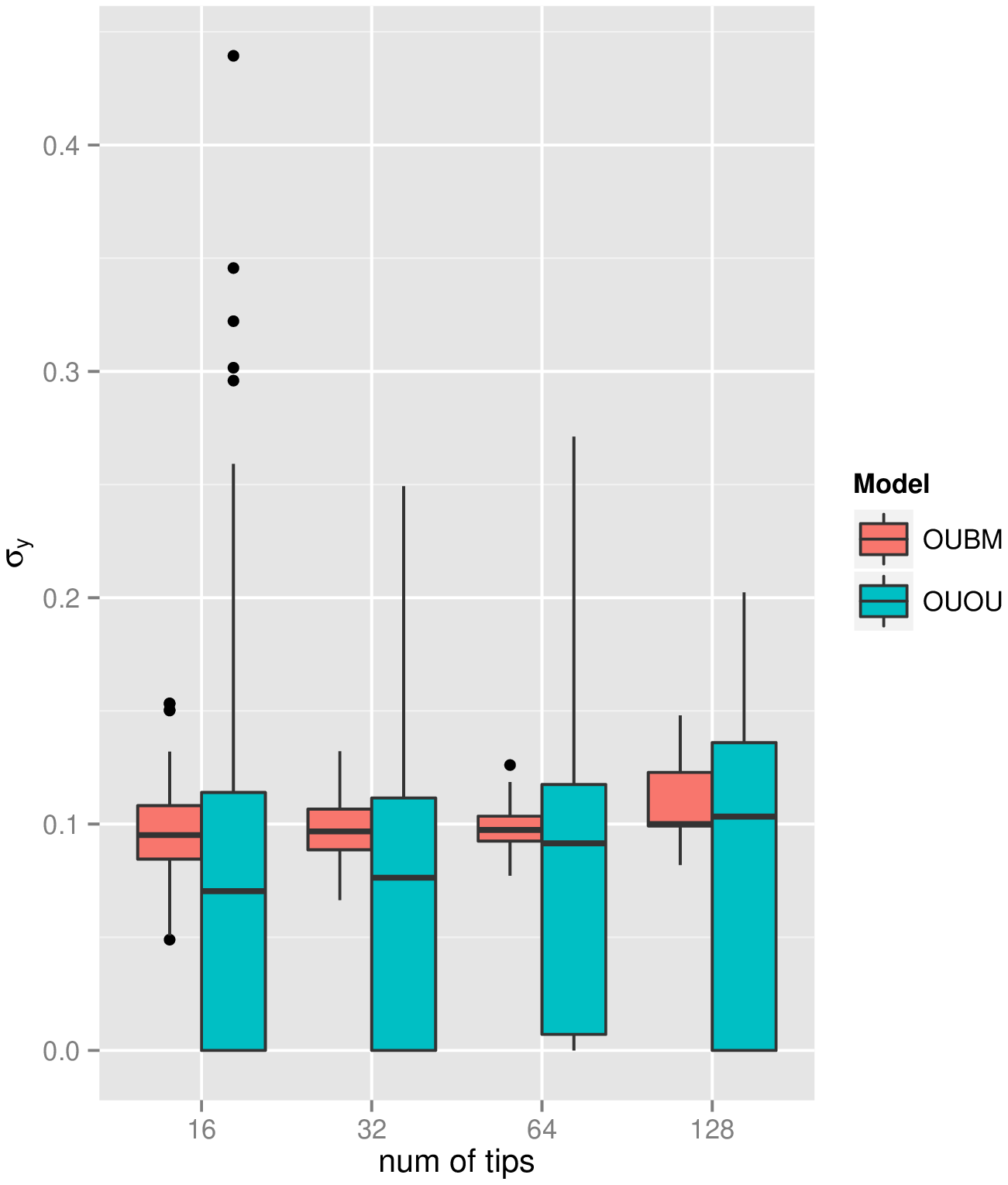}
	\end{center}
	\caption{Boxplot for $\sigma_y$.}
	\label{sigma_y}
\end{figure}

\begin{figure}
	\begin{center}
		\includegraphics[width=15cm,height=15cm]{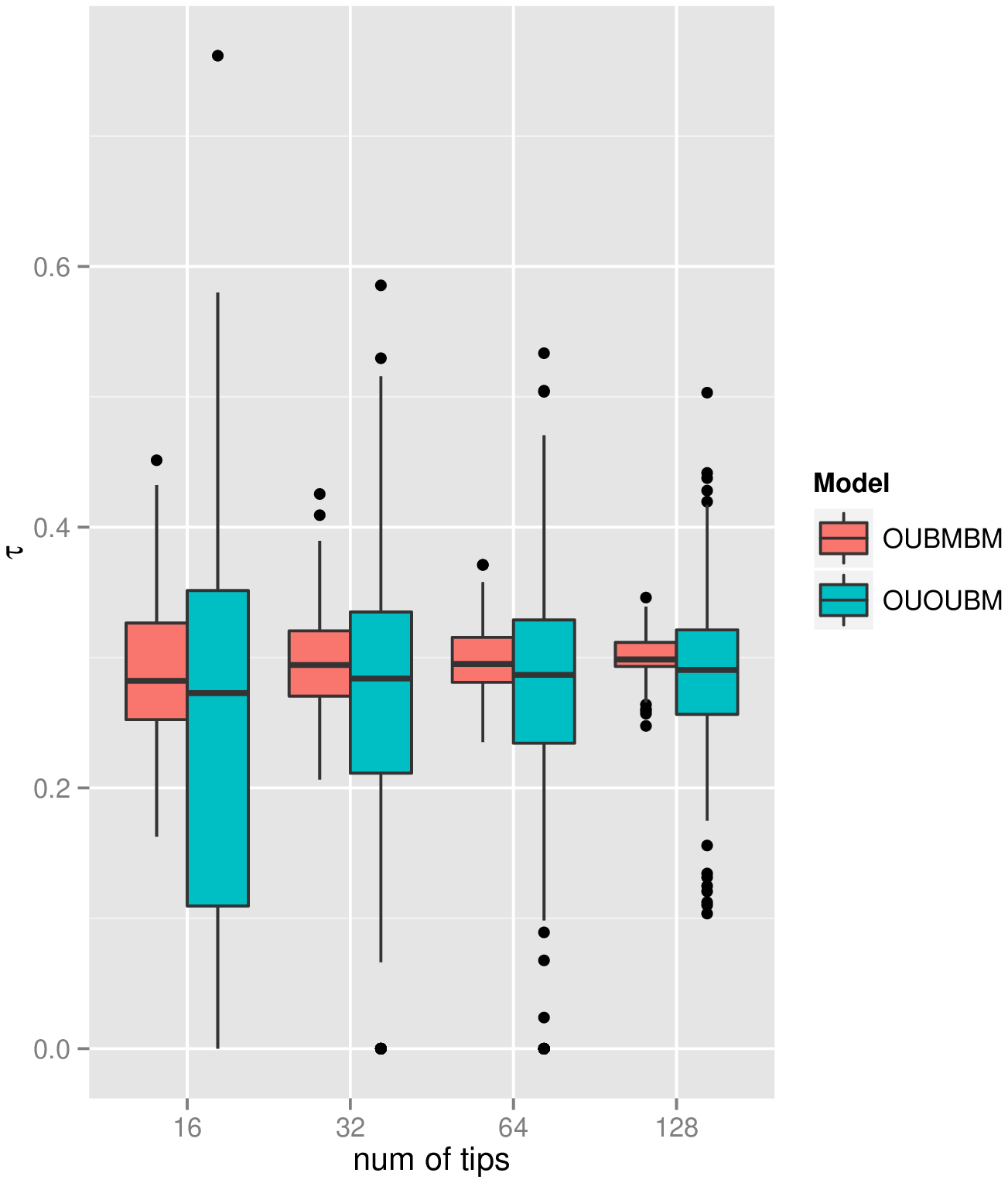}
	\end{center}
	\caption{Boxplot for $\tau$.}
	\label{tau}
\end{figure}

\end{document}